\newcommand{\la}{\ensuremath{\lambda}}

\newcommand{\be}{\begin{equation}}
\newcommand{\ee}{\end{equation}}
\newcommand{\del}{\partial}

\documentclass[12pt]{article}
\usepackage[dvips]{graphicx}
\usepackage{psfrag}
\begin{document}

\begin{titlepage}

\begin{flushright}
PUPT-1965 \\
RH-27-2000 \\
hep-th/0101080
\end{flushright}
\vspace{20 mm}

\begin{center}

{\Huge Zero-Mode Dynamics of String Webs}

\vspace{10mm}

{\large Paul Shocklee$^{a,b,}$\footnote{shocklee@princeton.edu} 
and L\'arus Thorlacius$^{b,}$\footnote{lth@hi.is}} 

\vspace{10mm}

a) Joseph Henry Laboratories\\ 
Princeton University \\
Princeton, New Jersey 08544 \\
USA

\vspace{5mm}

b) University of Iceland\\
Science Institute \\
Dunhaga 3, 107 Reykjavik \\ 
Iceland

\vspace{10mm}

{\large Abstract}
\end{center}

\noindent
At sufficiently low energy the dynamics of a string web is dominated 
by zero modes involving rigid motion of the internal strings.  The 
dimension of the associated moduli space equals the maximal number of 
internal faces in the web.  The generic web moduli space has boundaries 
and multiple branches, and for webs with three or more faces the geometry 
is curved.  Webs can also be studied in a lift to M-theory, where a 
string web is replaced by a membrane wrapped on a holomorphic curve in 
spacetime.  In this case the moduli space is complexified and admits 
a K\"ahler metric.

\vfil
\begin{flushleft}

\end{flushleft}

\end{titlepage}

\section{Introduction}

A few years ago it was realized that type IIB string theory not only 
contains an $SL(2,{\bf Z})$ multiplet of $(p,q)$ strings but also junctions 
where three or more of these strings come together \cite{schwarz,ofer}.
These junctions preserve a fraction of the IIB supersymmetry provided 
appropriate angles are chosen between the different $(p,q)$ strings 
\cite{dasgupta,sen}.  A collection of supersymmetric string junctions
can in turn be combined into a stable network of strings \cite{sen}, 
also referred to as a string web.  Such constructions are of intrinsic
interest as solitonic solutions of IIB string theory but they have also
found other applications.  They, for example, play a key role in 
realizing exceptional gauge symmetry in type IIB theory \cite{gaberdiel};
they represent ${1\over 4}$-BPS states in D=4, $\mathcal{N}=4$, $SU(N)$
supersymmetric Yang-Mills theory 
\cite{bergman, hashimoto, kawano, gauntlett2, monopoles};
and they enter into the AdS/CFT determination of the potential between
dyons at strong coupling in the $\mathcal{N}=4$ gauge theory \cite{minahan}.

String webs are dynamical objects with a rich spectrum of excitations.
Their dynamics has been studied in the small oscillation limit 
\cite{CandT, rey} and for generic string webs a set of rigid zero modes
has been identified \cite{kol,sethi}, which will dominate the dynamics at
sufficiently low energies.  In this paper we will primarily be interested
in these zero-mode motions.  We provide a general framework for studying 
the geometry of the associated moduli space, working out the explicit
metric for some examples.\footnote{The low-energy dynamics of string 
lattices has been recently discussed in \cite{sasakura2}.  In the present
paper, we are instead concerned with moduli of finite webs but the two
systems are related.}  

Supersymmetric string junctions and string webs may also be represented
in a lift to M-theory as a network of supermembranes holomorphically 
embedded into $\mathbf{R}^{1,8} \times \mathbf{T}^2$ where the integer 
charges of each $(p,q)$ string are realized as winding numbers of the 
corresponding membrane segment around the cycles of the 
$\mathbf{T}^2$~\cite{krogh,matsuo,sasakura}.
The membrane representation allows a systematic derivation of the 
modular dynamics.  The moduli space is complexified and has interesting 
geometric features which we explore alongside the more pedestrian string 
viewpoint.

The plan of the paper is as follows.  Section~2 reviews basic facts 
about string junctions and string webs, and is followed by a discussion
of the web moduli space and the dynamics of zero modes in Section~3.
We explicitly work out the metric on moduli space for some simple 
examples and give arguments about boundaries and multiple branches in
the general case.  Section~4 describes the lift to M-theory and our
concluding remarks are in Section~5.  In an appendix we prove that the
moduli space of any string web with only two internal faces is flat.

In this paper, we treat only bosonic zero-modes.  String webs are
supersymmetric systems and when the fermionic zero modes are included
the dynamics is governed by a supersymmetric effective theory which
we intend to present in a forthcoming paper \cite{pslt}.

\section{String junctions and string webs}

Three $(p,q)$ strings can come together to form a string junction, 
provided that the condition,
\begin{equation}
\sum_{i=1}^3 p_i = \sum_{i=1}^3 q_i = 0,
\end{equation}
is satisfied, which is necessary for charge conservation.  In order for 
this junction to be stable, the angles between the strings must be adjusted 
so that the net force on the vertex cancels.  
The tension of a $(p,q)$ string is given by
\be
T_{(p,q)} = T_{(1,0)}\, |p + q \tau|,
\ee
where $T_{(1,0)}$ is the fundamental string tension and $\tau$ is the 
axion-dilaton modulus of Type IIB theory.  Stability is guaranteed if
each $(p,q)$ string is oriented along the vector $(p + q \tau)$ in the 
complex plane \cite{dasgupta,sen}.  In this paper, we will for simplicity 
consider the self-dual point, $\tau = i$, so that fundamental $(1,0)$ strings
are oriented horizontally in all the figures while Dirichlet $(0,1)$ strings
are vertical.

One then constructs string webs by joining some number of these vertices 
together.  If all the strings in a given web lie in a single plane, with 
each string oriented parallel to its own charge vector, the web is stable 
and will in fact preserve 1/4 of the IIB supersymmetry \cite{sen}.  Of 
course, an overall rotation of the entire web does not affect its stability.
The charge conservation and stability conditions at a vertex are 
invariant under a change of sign of all the charges.  If we adopt the 
convention that strings are oriented outward from each vertex, there is
no ambiguity in charge assignments for strings that form external legs of
a web, but an internal string carries opposite charges with respect to 
each of the two vertices it ends at.

\begin{figure}
\includegraphics{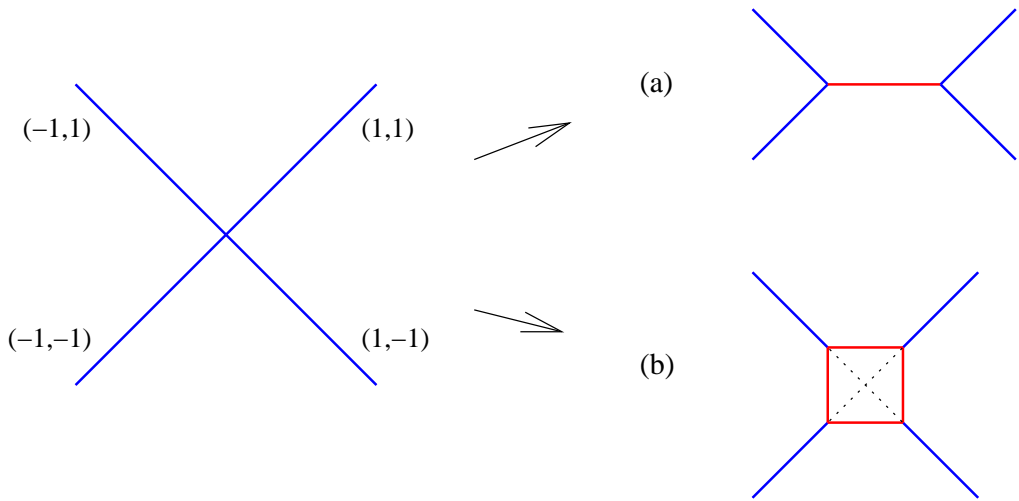}
\caption{Web deformations: (a) global deformation, (b) local deformation.} 
\label{fig:moduli}
\end{figure}

Certain supersymmetric string webs can be deformed without violating the 
local charge and stability conditions at any of the vertices.  These are 
either global deformations that move the external strings as in  
Figure~\ref{fig:moduli}a, or local deformations that only move internal strings
but leave the external ones unchanged as in Figure~\ref{fig:moduli}b. 
Note that $n$-point junctions with $n > 3$, as in Figure~\ref{fig:moduli}, 
are degenerate and can always be resolved into a web with only 3-point 
junctions by local and/or global deformations.

The total energy of a string web is preserved under local deformations.  
In other words, the total rest energy of the new strings that 
are created when a degenerate junction is ``blown up'' as in 
Figure~\ref{fig:moduli}b precisely equals the total rest energy of the 
string segments that have been removed from the original web in the 
process (shown as dotted lines in the figure).  Global deformations, on 
the other hand, involve shifting infinitely long segments of string and 
therefore do not correspond to physical motions of a string web
but rather to changing its defining parameters.  In other words, the
worldsheet zero-modes that correspond to global deformations are not
normalizable.
In applications to D=4 supersymmetric gauge theory the external strings 
end on 3-branes, which extend in directions orthogonal to the plane of 
the web but do not move within that plane, so that in this case the web 
has no global deformations \cite{bergman}.  In the present paper we will 
focus on local deformations and the moduli that parametrize them.

\section{Grid diagrams and zero-mode dynamics}

\begin{figure}
\psfrag{l}{$\ell$}
\includegraphics{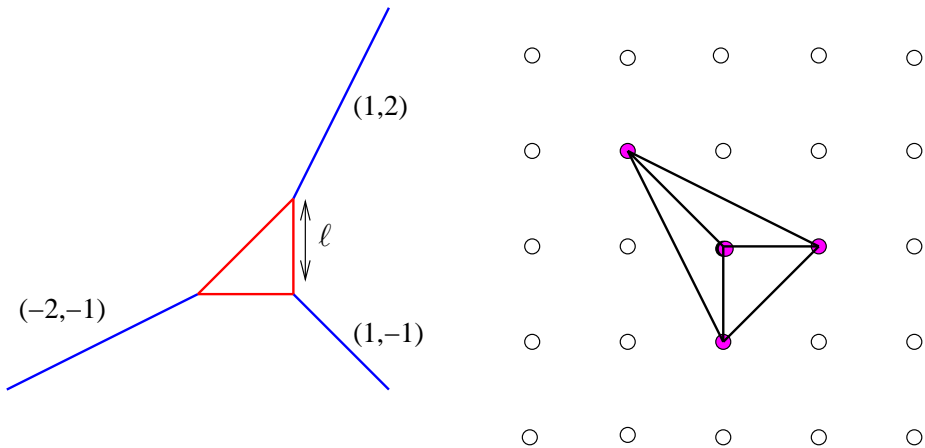}
\caption{A web with a single zero mode and its grid diagram.} 
\label{fig:one}
\end{figure}

\begin{figure}
\begin{center}
\psfrag{y=l1}{$y=\ell_1$}
\psfrag{x=-l2}{$x=-\ell_2$}
\includegraphics{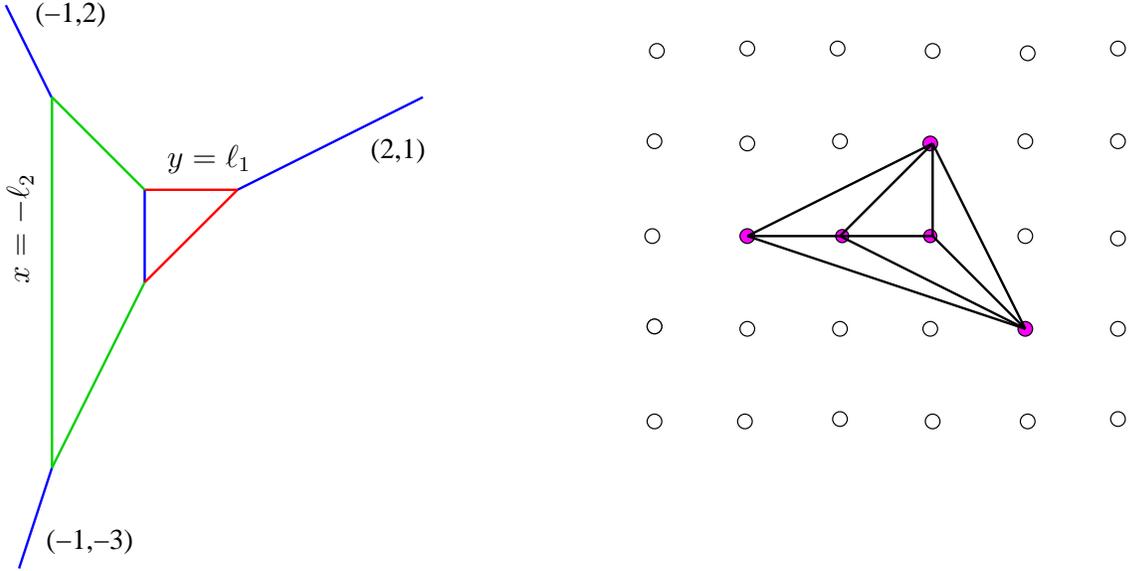}
\caption{A web with two zero modes and its grid diagram.} 
\label{fig:two}
\end{center}
\end{figure}

The number of local moduli of a given string web is governed by the 
$(p,q)$ charges carried by the external legs of the web and is equal to
the maximal number of internal faces in the web.  It can be read off 
from a ``grid diagram'', which is a dual description of the 
web \cite{bergman,kol,sethi}.  The rules for drawing grid diagrams are 
as follows (Figures~\ref{fig:one} and~\ref{fig:two} provide examples):
\begin{enumerate}
\item The diagram is drawn on a two-dimensional, integer, square lattice.
\item A $(p,q)$ string is represented by a $(-q,p)$ lattice vector.
\item Starting at an arbitrary lattice point, the vectors representing
the external strings are drawn in cyclic order, so that they form a
convex polygon.
\end{enumerate}
The number of local moduli is then given by the number of lattice points 
enclosed by the external polygon in the grid diagram.  The grid diagram 
in Figure~\ref{fig:one} has only one internal lattice point and thus the 
corresponding web has a single local modulus, which can, for example, be 
taken as the length $\ell$ of the vertical internal string.  The web in 
Figure~\ref{fig:two}, on the other hand, has two local moduli.

\par
The local moduli $\ell_i$ parametrize a set of web configurations that are
degenerate in total rest energy of string.  However, the total length of 
string involved in a rigid zero-mode motion on the moduli space clearly
depends on the value of the moduli.  In other words, the kinetic energy of
zero-mode motion is a function of the location on moduli space, which in 
turn defines a dynamical metric on the moduli space in the usual manner,
\begin{equation}
S = {1\over 2} \int dt\, g_{ij}(\ell) \,\dot{\ell}_i \,\dot{\ell}_j \,.
\end{equation}

As a simple example, let us consider the web in Figure~\ref{fig:one}.
This web has only a single modulus which we can take to be the length
$\ell$ of the vertical internal string.  A change in $\ell$ leads to 
transverse motion of each internal string in addition to changing its
length.  The total mass of string that moves is proportional to $\ell$
with a coefficient that depends on the tension of the individual strings.
Adding up the contribution from the three internal strings leads to the
following effective zero-mode action, 
\be
S = \frac{T_{(1,0)}}{6} \int dt~\ell \,\dot{\ell}^2 \,.
\label{eq:effact}
\ee
This is of course only the bosonic part.  We'll discuss fermions and
supersymmetry in~\cite{pslt}.
The change of variables $u=\alpha \ell^{3/2}$, with
$\alpha = 2 \sqrt{T_{(1,0)}}/3\sqrt{3}$, takes the system to the standard
one-dimensional free particle,
\be
S = \frac{1}{2} \int dt~\dot{u}^2 \,,
\ee 
reflecting the fact that any one-dimensional metric is flat.

Now consider the more complicated web of Figure~\ref{fig:two}, where the 
moduli space is two-dimensional.  We parametrize the moduli space by 
$(\ell_1,\ell_2)$ as follows:  Let the web lie in the xy-plane such that the 
external strings define the straight lines $y={1\over 2}x$, $y=-2x$, and 
$y=3x$.  Then let the horizontal (1,0) string be at $y=\ell_1$ and the
leftmost vertical (0,1) string at $x=-\ell_2$, as shown in 
Figure~\ref{fig:two}.  This is enough to uniquely determine the web 
configuration and is a convenient parametrization to compare with the 
membrane picture which we develop below.  Adding up contributions from the 
six internal strings one finds 
\begin{equation} \label{eq:2metric}
S = {T_{(1,0)}\over 2} \int dt\,\left [
3(3\ell_1-\ell_2)\dot{\ell_1}^2
-2(3\ell_1-\ell_2)\dot{\ell_1}\dot{\ell_2}
+(\ell_1+8\ell_2)\dot{\ell_2}^2 \right ] \,.
\end{equation}   
At first sight the moduli space metric appears non-trivial, but by defining
$\tilde{\ell_1}=\ell_1-{1\over 3}\ell_2$, $\tilde{\ell_2}=\ell_2$, followed 
by $\hat{\ell_1}=\alpha_1 \tilde{\ell_1}^{3/2}$,
$\hat{\ell_2}=\alpha_2 \tilde{\ell_2}^{3/2}$,
with appropriate constants $\alpha_1,\alpha_2$,
we once again arrive at a manifestly flat metric.  The geometry is flat
but the moduli space has boundaries; in terms of our parameters, we have 
the constraints $0 \le \ell_1 \le 2\ell_2$ and $0 \le \ell_2 \le 3\ell_1$.

\begin{figure} 
\label{fig:four}
\begin{center}
\includegraphics{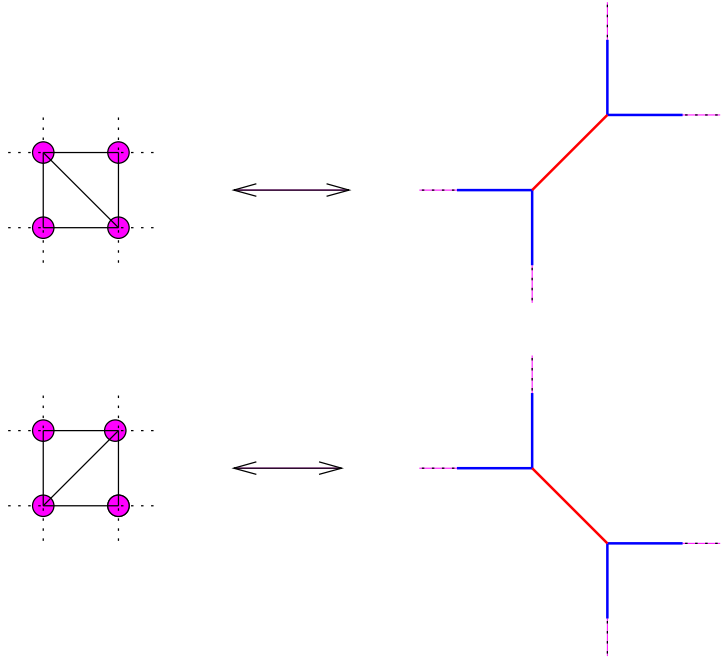}
\caption{Two branches in the moduli space related by a grid flip.} 
\end{center}
\end{figure}

Grid diagrams with only two internal lattice points are restricted in form 
and this places strong constraints on the geometry.  In fact, every 
two-parameter moduli space of string webs is flat.  A proof of this 
statement is given in the appendix.  By explicitly considering an example 
web with three internal faces we have, however, established that the metric 
generically has non-trivial curvature once the dimension of the moduli space
is three or more.  The flatness of the low-dimensional moduli spaces is 
presumably tied in with the high degree of supersymmetry of these systems.  

A general feature of string web moduli spaces is that the metric 
coefficients can be written as linear functions of the moduli,
\be
g_{ij}=\sum_{k=1}^I \alpha_{ijk} \, \ell_k \,,
\ee
where $I$ is the maximal number of internal faces in the web and the 
$\alpha_{ijk}$ are functions of the $(p,q)$ charges of the strings.

This method of adding ``by hand'' contributions to the kinetic energy from
individual internal strings gets tedious for more complicated webs and one
would like to find a more systematic approach.  As the number of internal
points in the grid diagram grows, the web moduli space not only becomes
curved but it is also branched.  To see this, consider a
grid diagram where the convex polygon representing the external strings
encloses a unit square in the grid lattice.  If all the string junctions
in the web are non-degenerate, {\it i.e.} connect only three strings,
then this part of the grid diagram takes one of the two forms shown in
Figure~4.  The corresponding string configurations differ
in the orientation of the string connecting two neighboring junctions as
shown in the figure.  When this structure is embedded in a larger string
web there will in general be local deformations that take the system from
one configuration to the other through a degenerate intermediate 
configuration containing a four-string junction.  These degenerate webs
correspond to a codimension one branching surface in the moduli space.  
For a web with many internal points in the grid diagram a large number of 
such ``grid flips'' is possible and the moduli space has multiple branches.

In the following section we describe how string webs may be represented
as wrapped membranes in M-theory.  In this approach the moduli space gets
complexified, the branching is smoothed out, and we find a systematic
prescription for computing the moduli space metric.

\section{Lift to M-theory}
In M-theory, string webs are represented by membranes wrapped on holomorphic 
curves embedded in $\mathbf{R}^{1,8} \times \mathbf{T}^2$ \cite{krogh,matsuo}.
Let us take the string web to lie in the $X^1X^2$-plane and parametrize the
torus by $X^9 \cong X^9 + 2\pi R$ and $X^{10} \cong X^{10} + 2\pi R$,
where $R$ is the compactification radius and we have set the torus modular
parameter (which is to be identified with the type IIB modulus) to 
$\tau = i$.  Now introduce complex coordinates 
\begin{equation}
Z^1 = X^1 + i X^9, \qquad Z^2 = X^3 + i X^{10}, 
\end{equation}
and then change to global variables
\begin{equation}
u = \exp(Z^1/R), \qquad v = \exp(Z^2/R).
\end{equation}
Consider a membrane embedded along the holomorphic curve defined by
\begin{equation}
u^qv^{-p} = \eta,
\end{equation}
with $\eta$ a non-zero complex constant and $p,q$ integers.
This curve represents a $(p,q)$ string.  Its projection onto the 
$X^1X^2$-plane is 
\begin{equation}
qX^1-pX^2 = R \log\vert\eta\vert \, ,
\end{equation} 
{\it i.e.} a straight line parallel to the $(p,q)$ charge vector whose 
position is fixed by $\eta$.  Meanwhile the projection onto the 
$\mathbf{T}^2$ shows that the
membrane is wrapped on the $(p,q)$ homology cycle.  
\par
A three-string junction consisting of a $(p_1,q_1)$, 
$(p_2,q_2)$, and a $(-p_1-p_2,-q_1-q_2)$ string is represented by a
single membrane embedded along the curve
\begin{equation} \label{eq:three}
\eta_1 u^{q_1}v^{-p_1} + \eta_2 u^{-q_2}v^{p_2}=\eta_3 \,,
\end{equation}
as is easily verified by considering the asymptotic behavior of the 
terms in (\ref{eq:three}) 
when $X^1$ and $X^2$ approach $\pm\infty$ in various combinations
\cite{krogh,matsuo}.
The complex coefficients $\eta_1$, $\eta_2$ and $\eta_3$ serve to fix the 
position of the junction.  
\par
The holomorphic equation for an arbitrary string web can be written down 
directly from the grid diagram \cite{kol,sethi}.  Once an origin is 
chosen for the lattice, the relation is given by 
\be\label{eq:membrane}
\sum_i^E \eta_i u^{m_i} v^{n_i} + \sum_j^I \la_j u^{m_j} v^{n_j} = 0\,,
\ee
where $(m_i,n_i)$ are the lattice coordinates of the $i$th vertex.  
Shifting the lattice origin amounts to multiplying all the terms by
a common factor and has no effect on the embedding.
The coefficients $\eta_i$ correspond to corner points of the convex 
polygon in the grid diagram while the $\la_j$ correspond to internal 
points and hence internal faces of the web.  Varying the $\eta_i$ and 
$\la_j$ amounts to global and local deformations of the web, respectively.
In this paper we are primarily interested in local deformation so we 
focus our attention on the $\la_j$, which parametrize the membrane 
moduli space.  The moduli space has $I$ complex dimensions, where $I$ is 
the number of internal faces of the web, or equivalently the genus of the 
embedded membrane in (\ref{eq:membrane}).
The string description is recovered in the limit where the handles of the
membrane become large compared to the compactification scale $R$, 
{\it i.e.} when $\log\vert\la_j\vert >>1$.

At first sight, the membrane moduli space shows no indication of the 
boundaries and branches of the string web moduli space.  For an 
infinitely extended web, each $\la_j$ ranges over the entire complex
plane.\footnote{Webs with the external strings ending on three-branes
are represented in M-theory by an open membrane with boundaries on 
five-branes that wrap the $\mathbf{T}^2$ and then the membrane moduli
space has finite volume.}
It turns out, however, that the membrane moduli space has curvature 
singularities precisely in regions that correspond to boundaries of
the string web moduli space.  These singularities only occur when one
or more of the moduli satisfy $\log\vert\la_j\vert\sim 1$, {\it i.e.}
the size of a handle is shrinking to the compactification scale $R$.
The locus of singularities is generally of (complex) codimension one
in the membrane moduli space.

\subsection{Zero-mode dynamics}
Let us first confirm that the $\la_j$ are moduli in the sense
that the rest energy of the membrane is independent of the values they
take.  This amounts to the membrane area being unchanged as the 
$\la_j$ are varied.  Since the membrane is described by a holomorphic
embedding it is a K\"ahler manifold of one complex dimension.  The 
area is then given by the integral of the K\"ahler form over 
the membrane, and we must show that this is constant.  In our complex 
structure, the K\"ahler form is given by the pullback of the 
following form to the membrane,
\begin{equation}
\omega = {i \over 2} (dZ^1 \wedge d\bar Z^1 
+ dZ^2 \wedge d\bar Z^2).
\end{equation}
In terms of $(u,v)$, this is 
\begin{equation}
\omega = {i \over 2} R^2 \left \{ {du \wedge d\bar u \over |u|^2}
+ {dv \wedge d\bar v \over |v|^2} \right \}.
\end{equation}
Integrating the pullback of this form over the membrane gives
\be
A = \frac{R^2}{2} \int du d\bar u~ 
\left ( \frac{1}{\vert u\vert^2} 
+ \left\vert\frac{1}{v} \frac{\del v}{\del u} \right\vert^2
\right ),
\ee
where we have chosen $u$ and $\bar u$ as independent variables and
$v$ and $\bar v$ are expressed as functions of $u$ and $\bar u$ 
through the embedding equation (\ref{eq:membrane}).  This integral
is divergent, but by introducing appropriate cutoffs one finds that
the divergence is associated with the infinite membrane pieces that
correspond to the infinite external strings \cite{krogh}.

Taking a partial derivative with respect to $\la_i$, and using the
fact that, for a holomorphic embedding, $\bar v$ is neither a function
of $u$ nor $\la_i$, we get 
\begin{eqnarray}
{\del A \over \del \la_i} &=& \frac{R^2}{2} \int d u d \bar u~\left (
-\frac{1}{v^2} \frac{\del v}{\del \la_i} \frac{\del v}{\del u} 
\frac{1}{\bar v} \frac{\del \bar v}{\del \bar u} 
+ \frac{1}{v} {\del^2 v \over \del \la_i \del u}
\frac{1}{\bar v} \frac{\del \bar v}{\del \bar u} \right )  \cr
&=& \frac{R^2}{2} \int d u d \bar u~\frac{\del}{\del u}
\left ( \frac{1}{v} \frac{\del v}{\del \la_i} \frac{1}{\bar v}
\frac{\del \bar v}{\del \bar u} \right ).
\end{eqnarray}
This could in principle get contributions from the asymptotic region
$|u| \rightarrow \infty$ and from points where the expression inside
the derivative diverges.  As mentioned above, the only possible 
singularities would come from regions associated with the external
strings and in those regions we have $\del v / \del \la_i 
\rightarrow 0$.  This, in turn, follows from the fact
that the asymptotic behavior of the membrane embedding in both $u$ and
$v$ is governed by the $\eta_i$ terms in (\ref{eq:membrane}).  
We conclude that the area is invariant under
local deformations and the $\la_j$ indeed are moduli. 

Let us consider now the dynamics of the corresponding zero modes.  
In this paper, we are only considering bosonic fields so the dynamics is 
governed by the Nambu-Goto action for the membrane,
\begin{equation}
S_{NG} = -T_{M2} \int d^3\sigma~\sqrt{\det{\gamma_{\alpha \beta}}},
\end{equation}
where
\begin{equation}\label{eq:indmetric}
\gamma_{\alpha \beta} = \del_\alpha X^\mu \del_\beta X_\mu,
\end{equation}
is the induced metric on the membrane worldvolume.
\par
We'll use a static gauge 
\begin{equation}
\sigma^0 = t,\qquad \sigma^1 = u,\qquad \sigma^2= \bar u,
\end{equation}
where $t\equiv X^0$ is embedding time, $u,\bar u$ are defined as before and 
$v,\bar v$ are obtained from the embedding equation (\ref{eq:membrane}).
The membrane is embedded into the $X^1,X^2,X^9,X^{10}$ 
directions only and we have $X^3 = \ldots  X^8 = 0$ 
for the remaining coordinates in (\ref{eq:indmetric}).

We now make the usual moduli space approximation, where we assume that
no oscillation modes of the membrane are excited and that time 
dependence of the membrane embedding enters only through the moduli:
\be
v = v(u;\la_j(t)) \,.
\ee
Expanding the action for slowly varying moduli, we find
\be\label{eq:memact}
S_{NG} = T_{M2} \int dt \left( -A 
+ \frac{1}{2} R^4 g_{i \bar j} {\dot \la}_i
{\dot{\bar \la}}_j + {\cal O}({\dot \la}^4) \right) \,,
\ee
where $A$ is the membrane area and the metric on moduli space is
given by
\begin{equation} \label{eq:metric}
g_{i \bar j} = {1\over 2}\int du d\bar u~\frac{1}{|uv|^2} 
\frac{\partial v}{\partial \la_i}
\frac{\partial \bar v}{\partial \bar \la_j}.
\end{equation} 
This is manifestly a K\"ahler metric, 
$g_{i \bar j} = \frac{\partial}{\partial \la_i} 
\frac{\partial}{\partial \bar \lambda_j} {\cal K}$, 
with the K\"ahler potential reducing to a particularly simple formal
expression in terms of $Z^2(Z^1)$,
\begin{equation}
{\cal K} = \frac{1}{2R^4} \int d Z^1 d \bar Z^1 |Z^2|^2 \,. 
\end{equation}

\subsection{Explicit examples}
Let us now apply this formalism to concrete examples.  This reveals
some of the geometry involved and allows us to test the general formula
(\ref{eq:metric}) by checking explicit answers against the corresponding 
string web results.

Consider the junction in Figure~\ref{fig:one}.
It has one local deformation, and is described by the curve
\begin{equation} \label{eq:curve}
u + u^2 v + v^2 + \la u v = 0.
\end{equation}
Using equation (\ref{eq:metric}), the metric on this moduli 
space is given by
\begin{equation}
g_{\la \bar \lambda} = \frac{1}{2}\int du d\bar u~\frac{1}{|u^2 
+ 2 v + \lambda u|^2}.
\end{equation}
Since the curve (\ref{eq:curve}) is quadratic in $v$, we can solve for 
$v$, leaving us with the following explicit integral expression for   
the metric,
\begin{equation} \label{eq:explicit}
g_{\la \bar \lambda} = \frac{1}{2} \int du d\bar u 
{1 \over |u(u(u + \lambda)^2 - 4)|}.
\end{equation}
To make contact with the string web we consider the
limit of large $\vert\la\vert$.  In this limit, the internal face
of the junction becomes large compared to the M-theory compactification 
scale $R$.  By examining the projection of the curve (\ref{eq:curve}) 
onto the $X^1X^2$-plane one identifies the string web modulus as 
$\ell=3R\log\vert\la\vert$.  We then write the complex membrane
modulus as
\be
\la = \exp ({l\over 3 R}+i\theta) \,,
\ee
and consider the change of variables from $(\la,\bar\la)$ to 
$(l,\theta)$.  In the limit $\ell >>R$ one finds that the 
$g_{\theta\theta}$ term in the membrane kinetic energy may be dropped
in comparison to the $g_{\ell\ell}$ term, effectively reducing the
dimension of the moduli space from one complex to one real dimension.
Furthermore, a numerical evaluation of the integral (\ref{eq:explicit})
shows that the kinetic part of the membrane action (\ref{eq:memact})
reduces to the string web action (\ref{eq:effact}), including the correct
normalization, when we identify the fundamental string tension as
$T_{(1,0)}=2\pi R T_{M2}$.

The integrand in (\ref{eq:explicit}) is singular whenever the denominator 
has a zero.  These singularities are integrable, unless there is a 
double zero, and then the metric itself is singular.  
In our example, this occurs when 
$\la = -3,~3e^{\pi i/3},~3e^{-\pi i/3}$, and one easily checks that 
these are curvature singularities.  They all occur at a distance of 
order $R$ away from the origin in the $X^1X^2$-plane, which corresponds
to the $\ell=0$ boundary of the string web moduli space, 
in the $\ell>>R$ string limit. 

Let us also consider a two-dimensional example corresponding to the 
string web in Figure~\ref{fig:two}.  The algebraic 
curve for this junction has two deformation parameters, which we can call 
$\la_1$ and $\la_2$.  The curve is given by the equation
\begin{equation} \label{eq:twomode}
u^2 v^2 - v - u^3 + \la_1 u^2v + \la_2 u v = 0.
\end{equation}
The membrane parameters are related to those in the string picture by 
$\la_i=\exp{(\ell_i/R)}$.
This curve is also quadratic in $v$, so we can again solve 
for $v$ and express the metric on the moduli space explicitly in terms
of integrals,
\begin{equation}
g_{i \bar j} \dot \la_i \dot{\bar \lambda}_j 
= \frac{1}{2} \int d u d \bar u 
{1 \over |(1 - \la_1 u^2 - \la_2 u)^2 + 4 u^5|} 
(|u|^2 |\dot \la_1|^2 + |\dot \la_2|^2
+ u \dot \la_1 \dot{\bar \la_2} + \bar u \dot{\bar \la_1} \dot \la_2).
\end{equation}
In this case, the moduli space has two complex dimensions, and it has
singularities that lie on complex lines, corresponding to the boundaries
of the moduli space in the string limit.  We have evaluated these integrals
numerically in the limit of large $\vert\la_i\vert$,
and we find that they reproduce the kinetic energy (\ref{eq:2metric}) of 
the string web moduli.  It is interesting to note that the two-dimensional
moduli space of the string web is everywhere flat, but may nevertheless be
obtained as a slice through a non-trivial geometry, of complex dimension 
two, that has curvature singularities.

\section{Conclusions}

We have presented two methods for computing the metric on moduli space
for a string web.  One is to add up the kinetic energies of all the strings
that are involved in the zero-mode motion.  This method provides some 
physical insight but the necessary plane geometry gets tedious as the 
number of strings in the web grows.  The moduli space of a string web 
with one or two internal faces is flat but has boundaries.  The
higher-dimensional moduli spaces have boundaries and also non-trivial 
curvature and topology, with multiple branches.

The second method involves lifting to M-theory where the string web is 
represented as a membrane holomorphically embedded in 
$\mathbf{R}^{1,8} \times \mathbf{T}^2$.  The resulting moduli space is
complex and admits a K\"ahler metric, which is directly obtained, up to
quadrature, from the analytic curve defining the membrane embedding.
Numerical evaluation of the required integrals in the appropriate limit
gives results consistent with the previous method.  One can obtain useful
information by analytic methods without explicitly evaluating the
metric.  Although the complex moduli space has no boundaries, one instead 
finds curvature singularities in regions where the corresponding string 
web has internal faces shrinking to zero size.  We expect regions of 
strong curvature to be repulsive in the effective quantum dynamics on 
the moduli space and thus to mimick the effect of boundaries.  The 
actual singularities should perhaps not be taken too literally since 
they occur in regions of moduli space where the size of a membrane 
handle approaches the M-theory compactification scale $R$, and the 
classical supergravity description of our membranes is breaking down.
One expects additional light degrees of freedom to emerge in the 
theory at that point, which resolve the singularity.  Such degrees of
freedom are readily apparent in the string web picture, where 
fundamental strings, attached to the solitonic strings of the web and
stretched across a web face, become massless when the face shrinks to
zero size, and need to be included in the low-energy dynamics.

Our description of the zero-mode dynamics of superstring webs in this 
paper is missing some crucial ingredients.  In particular, we need to
incorporate fermion zero modes in order to explore the supersymmetry
of the low-energy dynamics.  Our methods can be generalized to include
fermions by considering supermembranes embedded in the superspace of
$D=11$ supergravity \cite{pslt}.

\noindent
\underline{Acknowledgments:}\hfill\break
We would like to thank B. Greene, B. Kol, M. Krogh, and J. Wang for 
helpful discussions.  This work was supported in part by an NSF 
Graduate Student Fellowship and by grants from the Icelandic Research
Council and the University of Iceland Research Fund.

\vskip\the\baselineskip

\appendix
\noindent{\Large\bf Appendix}

\begin{figure} \label{fig:genweb}
\begin{center}
\psfrag{(p1,q1)}{\small $(p_1,q_1)$}
\psfrag{(p2,q2)}{\small $(p_2,q_2)$}
\psfrag{(r1,s1)}{\small $(r_1,s_1)$}
\psfrag{(r2,s2)}{\small $(r_2,s_2)$}
\includegraphics{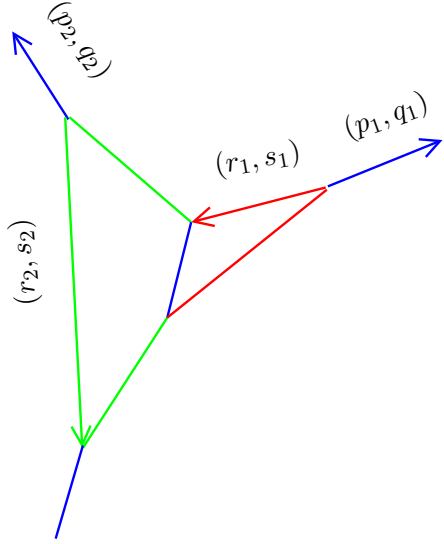}
\caption{A generic string web with two zero modes.} 
\end{center}
\end{figure}

\vskip 10pt
\noindent
A generic string web with three external
strings and two internal faces is shown in Figure~\ref{fig:genweb}.  
It is a general feature of such webs that one of the faces meets only one 
of the external strings, to which we have assigned the charges $(p_1,q_1)$, 
while the other face connects to the two remaining external strings.  
By an overall rotation we can always make the $(p_1,q_1)$ string 
lie in the upper-right-hand quadrant as in the figure.  We have indicated
the charges and orientation of a few of the strings in the web.  The
charge assignments for the remaining strings follow uniquely from charge 
conservation at the vertices.

We have, for convenience, arranged the global moduli to be such that when 
both faces shrink to zero size, the three external strings meet at a point, 
which we take as our origin in the xy-plane of the web.  The external 
strings then lie on the straight lines $p_1y=q_1x$, $p_2y=q_2x$, and
$(p_1+p_2)y=(q_1+q_2)x$.  We introduce the two moduli $\ell_1,\ell_2$ by
parametrizing the two labelled internal strings in Figure~\ref{fig:genweb}
as follows: $r_1y=s_1x-\ell_1$ and $r_2y=s_2x-\ell_2$.  Note that the web
in Figure~\ref{fig:two} is a special case of this construction with
$p_1=2$, $q_1=1$, $p_2=-1$, $q_2=2$, $r_1=-1$, $s_1=0$, $r_2=0$, and
$s_2=-1$.

It is now a straightforward, if tedious, exercise to add up the kinetic 
energies of the six internal strings when $\ell_1$ and $\ell_2$ depend 
slowly on time.  The result is a rather messy expression constructed
out of $SL(2,Z)$ invariants of the form $p_i q_j -p_j q_i$ involving
various combinations of the string charges.  Rather than presenting all
the metric components here and establishing flatness by direct 
computation, we use the following shortcut.  As mentioned in the main
text, generic metric components are linear functions of the moduli,
\be
g_{ij}= a_{ij}\ell_1 + b_{ij}\ell_2 \,,
\ee
with $i,j=1,2$.  The coefficients $a_{ij}$, $b_{ij}$ are functions of 
$SL(2,Z)$ invariants.  This metric is certainly flat if it can be 
brought into the form 
\be
\label{eq:flatmetric}
[g_{\tilde{i}\tilde{j}}] = \left[
\begin{array}{cc}
\alpha_1\tilde{\ell_1} & 0 \\
0 & \alpha_2\tilde{\ell_2}
\end{array}
\right] \,,
\ee
by a linear transformation of the moduli,
\be
\left( \begin{array}{c}
\tilde{\ell_1} \\
\tilde{\ell_2}
\end{array} \right)
= \left[ \begin{array}{cc}
A & B \\
C & D
\end{array} \right]\,
\left( \begin{array}{c}
\ell_1 \\
\ell_2
\end{array} \right) \,.
\ee
The metric coefficients satisfy the following relations:
\begin{eqnarray}
a_{11}=A^3\alpha_1+C^3\alpha_2, \phantom{BB} 
& b_{11}=A^2B\alpha_1+C^2D\alpha_2, \\
a_{12}=A^2B\alpha_1+C^2D\alpha_2,  
& b_{12}=AB^2\alpha_1+CD^2\alpha_2, \\
a_{11}=AB^2\alpha_1+CD^2\alpha_2,  
& b_{11}=B^3\alpha_1+D^3\alpha_2. \phantom{BB}
\end{eqnarray}
We note from this that a sufficient condition for flatness is to have
the relations $b_{11}=a_{12}$ and $b_{12}=a_{22}$ satisfied by our
original metric, while $a_{11}$ and $b_{22}$ are unrestricted.  By 
explicit calculation we find
\begin{eqnarray}
b_{11}=a_{12}=\bigl(q_1(p_2+r_2)-p_1(q_2+s_2)\bigr)/D \,,  \\
b_{12}=a_{22}=\bigl(s_1p_1-r_1q_1\bigr)/D \,, \phantom{MMMNIII}
\end{eqnarray}
with $D=\bigl(s_1(p_2+r_2)-r_1(q_2+s_2)\bigr)
\bigl((p_1+r_1)(q_1+q_2+s_2)-(q_1+s_1)(p_1+p_2+r_2)\bigr)$, showing
that the moduli space of a generic web with two internal faces is
indeed flat.


\begin{thebibliography}{99}
\bibitem{schwarz} 
J.H.~Schwarz, ``An $SL(2,{\bf Z})$ Multiplet of Type IIB Superstrings'',
{\em Phys. Lett.} {\bf B360} (1995) 13, {\tt hep-th/9508143}; ``Lectures
on superstring and M-theory dualities'',
{\em Nucl. Phys. B} (Proc. Suppl.) {\bf 55B} (1997) 1, {\tt hep-th/9607201}.
\bibitem{ofer}
O.~Aharony, J.~Sonnenschein, and S.~Yankielowicz,
``Interactions of strings and D-branes from M-theory'', {\em Nucl. Phys.} 
{\bf B474} (1996) 309, {\tt hep-th/9603009}.
\bibitem{dasgupta} K.~Dasgupta and S.~Mukhi, ``BPS nature of 3-string 
junctions'', {\em Phys. Lett.} {\bf B423}, 261 (1998), {\tt hep-th/9711094}.
\bibitem{sen} A.~Sen, ``String network'', {\em JHEP} {\bf 03} (1998) 005, 
{\tt hep-th/9711130}.
\bibitem{gaberdiel} M.R.~Gaberdiel and B.~Zweibach, ``Exceptional groups 
from open strings'', {\em Nucl. Phys.} {\bf B518} (1998) 151, 
{\tt hep-th/9709013}; M.R.~Gaberdiel, T.~Hauer, and B.~Zwiebach, ``Open 
string-string junction transitions'', {\em Nucl. Phys.} {\bf B525} 
(1998) 117, {\tt hep-th/9801205}.
\bibitem{bergman} O.~Bergman, ``Three-pronged strings and $1/4$ BPS states 
in $\mathcal{N}=4$ super-Yang-Mills theory'', {\em Nucl. Phys.} {\bf B525}
(1998) 104, {\tt hep-th/9712211};
O.~Bergman and B.~Kol, ``String webs and $1/4$ BPS monopoles'',
{\em Nucl. Phys.} {\bf B536} (1998) 149, {\tt hep-th/9804160}.
\bibitem{hashimoto} K.~Hashimoto, H.~Hata, and N.~Sasakura,
``3-String Junction and BPS Saturated Solutions in SU(3) Supersymmetric
Yang-Mills Theory'', {\em Phys. Lett.} {\bf B431} (1998) 303, 
{\tt hep-th/9803127}; ``Multi-Pronged Strings and BPS Saturated Solutions 
in SU(N) Supersymmetric Yang-Mills Theory'', {\em Nucl. Phys.}
{\bf B535} (1998) 83, {\tt hep-th/9804164}.
\bibitem{kawano} T.~Kawano and K.~Okuyama, ``String network and $1/4$ BPS 
states in $\mathcal{N}=4$ $SU(n)$ supersymmetric Yang-Mills theory'', 
{\em Phys. Lett.} {\bf B432} (1998) 338, {\tt hep-th/9804139}. 
\bibitem{gauntlett2} J.~Gauntlett, C.~Koehl, D.~Mateos, P.~Townsend, and 
M.~Zamaklar, ``Finite energy Dirac-Born-Infeld monopoles and string 
junctions'', {\em Phys. Rev.} {\bf D60} (1999) 045004, {\tt hep-th/9903156}.
\bibitem{monopoles} B.~Kol and M.~Kroyter, ``On the spatial structure of 
monopoles'', {\tt hep-th/0002118}.
\bibitem{minahan} J.A.~Minahan, ``Quark - monopole potentials in large $N$
super Yang-Mills'',  {\em Adv. Theor. Math. Phys.} {\bf 2} (1998) 559, 
{\tt hep-th/9803111}. 
\bibitem{CandT} C.G.~Callan and L.~Thorlacius, ``Worldsheet dynamics of 
string junctions'', {\em Nucl. Phys.} {\bf B534} (1998) 121, 
{\tt hep-th/9803097}.
\bibitem{rey} S.-J.~Rey and J.-T.~Yee, ``BPS dynamics of triple $(p,q)$ 
string junction'', {\em Nucl. Phys.} {\bf B526} (1998) 229, 
{\tt hep-th/9711202}.  
\bibitem{kol} O.~Aharony, A.~Hanany, and B.~Kol, ``Webs of $(p,q)$ 5-branes, 
five-dimensional field theories, and grid diagrams'', {\em JHEP} {\bf 01} 
(1998) 002, {\tt hep-th/9710116}.
\bibitem{sethi} A.~Mikhailov, N.~Nekrasov, and S.~Sethi, ``Geometric
realizations of BPS states in ${\cal N} = 2$ theories'',
{\em Nucl. Phys.} {\bf B531} (1998) 345, {\tt hep-th/9803142}.
\bibitem{sasakura2} N.~Sasakura, ``Low-energy propagation modes on string
network'', {\tt hep-th/0012270}.
\bibitem{krogh} M.~Krogh and S.~Lee, ``String network from M-theory'', 
{\em Nucl. Phys.} {\bf B516} (1998) 241, {\tt hep-th/9712050}.
\bibitem{matsuo} Y.~Matsuo and K.~Okuyama, ``BPS condition of string junction
from M-theory'', {\em Phys. Lett.} {\bf B426} (1998) 294, {\tt hep-th/9712070}.
\bibitem{sasakura} N.~Sasakura and S.~Sugimoto, ``M-theory description of 
$1/4$ BPS states in ${\cal N} = 4$ supersymmetric Yang-Mills theory'', 
{\em Prog. Theor. Phys.} {\bf 101} (1999) 749, {\tt hep-th/9811087}.
\bibitem{pslt} P.~Shocklee and L.~Thorlacius, in preparation.
\end{thebibliography}
\end{document}